\documentclass[12pt, onecolumn, journal]{IEEEtran}

\usepackage{amsmath,amssymb,amscd}
\usepackage{graphicx,cite}
\usepackage{algorithm}
\usepackage{algpseudocode}

\begin{document}
\title{Optimal Fronthaul Quantization for Cloud Radio Positioning}
\author{\IEEEauthorblockN{Seongah Jeong, Osvaldo Simeone, Alexander Haimovich and Joonhyuk Kang}
\thanks{Seongah Jeong and Joonhyuk Kang are with the Department of Electrical Engineering, Korea Advanced Institute of Science and Technology (KAIST), Daejeon, South Korea (Email: seongah@kaist.ac.kr and jhkang@ee.kaist.ac.kr).

Osvaldo Simeone and Alexander Haimovich are with the Center for Wireless Communications and Signal Processing Research (CWCSPR), ECE Department, New Jersey Institute of Technology (NJIT), Newark, NJ 07102, USA (Email: osvaldo.simeone@njit.edu and haimovic@njit.edu). 
}\\
}
\maketitle
\begin{abstract}
Wireless positioning systems that are implemented by means of a Cloud Radio Access Networks (C-RANs) may provide cost-effective solutions, particularly for indoor localization. In a C-RAN, the baseband processing, including localization, is carried out at a centralized control unit (CU) based on quantized baseband signals received from the RUs over finite-capacity fronthaul links. In this paper, the problem of maximizing the localization accuracy over fronthaul quantization/compression is formulated by adopting the Cram\'{e}r-Rao bound (CRB) on the localization accuracy as the performance metric of interest and information-theoretic bounds on the compression rate. The analysis explicitly accounts for the uncertainty of parameters at the CU via a robust, or worst-case, optimization formulation. The proposed algorithm leverages the Charnes-Cooper transformation and Difference-of-Convex (DC) programming, and is validated via numerical results.      
\end{abstract}

\begin{IEEEkeywords} Cloud Radio Access Networks (C-RANs), quantization, localization, Cram\'{e}r-Rao bound (CRB).  
\end{IEEEkeywords}

\section{Introduction}\label{sec:intro}
\begin{figure}[t]
\begin{center}
\includegraphics[width=9cm]{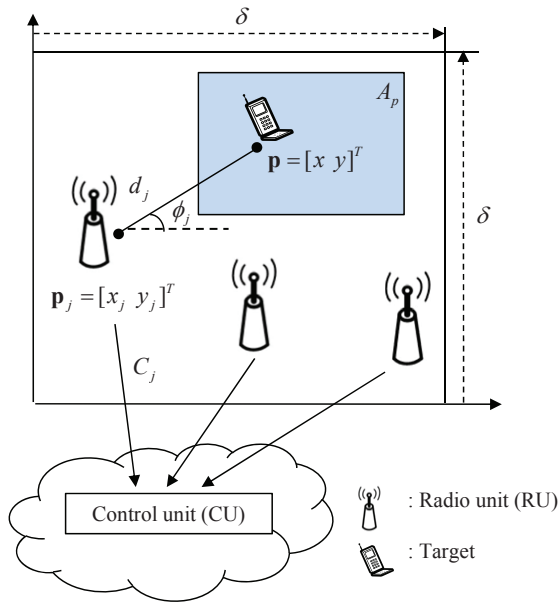}
\caption{Illustration of the considered Cloud Radio Positioning system, which consists of $N_r$ distributed single-antenna RUs, a single-antenna target, e.g., a mobile terminal, and a CU. The RUs are connected to the CU via finite-capacity fronthaul links. The target is known a priori to be in a given uncertainty region $A_p$.} \label{fig:sys}
\end{center}
\end{figure}

Cloud Radio Access Networks (C-RANs) provide a novel architecture for wireless cellular systems, whereby all baseband processing is migrated from the base stations (BSs) to a centralized control unit (CU). In the uplink of a C-RAN, the role of the BSs is hence reduced to that of radio units (RUs) that downconvert the received radio signals, which are then digitized and sent on fronthaul links to the CU. A key limitation of C-RANs is the finite capacity available on the fronthaul link connecting the RUs to the CU (see Fig. \ref{fig:sys}). This is dealt with via the implementation of compression strategies at the RUs that aim at reducing the bit rate produced by the digitized baseband signals \cite{ChinaMobile}. 

A key requirement in modern cellular system is location-awareness, which finds applications for security, disaster response, emergency relief and surveillance \cite{Gezici_05spmag}. In GPS-denied environments, positioning can be provided by wireless cellular networks, as recently mandated by the new FCC requirements on indoor positioning \cite{FCC15}. If the CU, or fusion center, has access to the signals received by the BSs, or more generally wireless sensors, it can perform localization by means of various methods based on the estimation of time of arrival (TOA) \cite{Li13TON, Shen14ACM, JSA15TVT}, time difference of arrival (TDOA) \cite{JSA15TVT}, angle of arrival (AOA) \cite{Gezici_05spmag} or received signal strength (RSS) \cite{Masazade12TSP, Ozdemir09TSP}. A localization techniques based on TDOA has been standardized by the 3rd Generation Partnership Project (3GPP) for use in the Long Term Evolution (LTE) systems\cite{EUTRA455}.

In this work, we study the problem of positioning in the C-RAN uplink system illustrated in Fig. \ref{fig:sys}, which we refer to as providing Cloud Radio Positioning. In this system, a key problem is that of designing fronthaul quantization and compression strategies to cope with fronthaul capacity limitations. In \cite{Masazade12TSP,Ozdemir09TSP}, RSS-based localization is considered based on nonuniform scalar quantizer. Unlike \cite{Masazade12TSP,Ozdemir09TSP}, here we assume a direct localization approach, which could be directly implemented in a C-RAN: the CU estimates the position of the target based on the (complex) baseband signals received by the RUs which are quantized and compressed for transmission on the fronthaul links. Direct localization is known to outperform indirect, or two-step, localization in which the estimate is based on parameters, such as TOAs, evaluated at the RUs \cite{Weiss04TSP}. Moreover, we account for the uncertainty of parameters, such as inter-node distance, angle and channel gains, at the CU via a robust, or worst-case, optimization formulation \cite{Ben09Book}, rather than the average performance criterion used in \cite{Masazade12TSP, Ozdemir09TSP}. We adopt the Cram\'{e}r-Rao bound (CRB) on the localization accuracy as the performance metric of interest and information-theoretic bounds on the compression rate. Specifically, after introducing the system model (Section \ref{sec:sm}), we propose an algorithm that solves the robust optimization problem at hand via the Charnes-Cooper transformation and Difference-of-Convex (DC) programming (Section \ref{sec:opt}). The algorithm is verified via numerical results (Section \ref{sec:numerical}). 
\section{System Model}\label{sec:sm}
We consider a Cloud Radio Positioning system consisting of $N_r$ distributed single-antenna RUs and a CU, whose goal is locating a single-antenna radio transmitter (see Fig. \ref{fig:sys}). The RUs may account for different types of infrastructure nodes such as macro/femto/pico BSs, relay stations or distributed antennas. The set of RUs is denoted as $\mathcal{N}_r= \{1, \dots, N_r\}$ and is placed within a $\delta \times \delta$ square region. The RUs are connected to the CU via finite-capacity fronthaul links. Based on the signals received on the fronthaul links from the RUs, the CU aims at locating a radio transmitter, e.g., a mobile station, whose position is $\pmb{p}=[x\,\,y]^T$ and is known a priori to lie in a given region $A_p$, which may be smaller than the overall square region. We will refer to the terminal to be located as the target. Each RU $j$ for $j \in \mathcal{N}_r$ is located at position $\pmb{p}_j=[x_j\,\,y_j]^T$ in the entire area, and the positions of all RUs are assumed to be known to the CU. The distance and angle between the target and RU $j$ are defined as $d_j=||\pmb{p}-\pmb{p}_j||$ and $\phi_j=\tan^{-1}((y-y_j)/(x-x_j))$, respectively. We assume $N_r \ge 3$ so that the target's position $\pmb{p}$ to avoid ambiguities (see, e.g., \cite{Li13TON, Shen14ACM, JSA15TVT}).

In order to enable the CU to locate the target, each RU $j$ downconverts the received signal to baseband, compresses and forwards it to the CU through the corresponding fronthaul link. The fronthaul link between RU $j$ and CU is assumed to have capacity $C_j$ (bits/s/Hz), where the normalization is with respect to the bandwidth of the signal transmitted by the target. Note that the fronthaul links can be either wireless, e.g., a microwave link, or wired, e.g., a coaxial cable or a fiber optics link. The impact of the fronthaul capacity limitations will be further discussed below.
\subsection{Signal Model}\label{sec:signal model}
We start by detailing the system model. The channel between the target and each RU $j$ is frequency-flat and is described by the impulse response $h_j(t)=g_j\delta(t-\tau_j)$, or, equivalently, by the transfer function $H_j(f)=g_j e^{-j2\pi f\tau_j}$. The parameter $\tau_j$ is the propagation delay of the path between the target and the RU $j$, which depends on the target's position as $\tau_j=||\pmb{p}-\pmb{p}_j||/c$ with $c$ being the propagation speed. Note that we assume the presence of a common time reference between the target and all the RU $j$ for all $j \in \mathcal{N}_r$. A time mismatch could be accounted for as in \cite{JSA15TVT} but this is not further pursued here. Also, the parameter $g_j$ models the path loss as $g_j=\alpha_j/d_j^\mu$ with $\mu$ being the path loss exponent and $\alpha_j$ being the independent channel fading coefficient for $j \in \mathcal{N}_r$, which is assumed to have power $\sigma_{\alpha_j}^2=E[|\alpha_j|^2]$ so that the power of the channel gain $g_j$ is $\sigma_{g_j}^2=\sigma_{\alpha_j}^2/d_j^{2\mu}$. We assume that each RU and the CU are informed only about the channel fading powers $\sigma_{\alpha_j}^2$ but not about the instantaneous values $\alpha_j$. These powers can be estimated, e.g., via RSS measurements.

The target transmits the upconverted version of a baseband signal $x(t)$ to the RUs. The signal $x(t)$ is assumed to be a training sequence known to all the nodes, and its Fourier transform and energy spectral density (ESD) are denoted by $X(f)$ and $S_x(f)=|X(f)|^2$, respectively. The baseband waveform received at the RU $j$ can be written as $y_j(t)=h_j(t)\ast x(t)+z_j(t)=g_jx(t-\tau_j)+z_j(t)$, where $z_j(t)$ is a stationary complex baseband Gaussian random process that represents the signal-independent disturbance, which includes the contribution of the noise and also of the interference from possible coexisting systems. The power spectral density of $z_j(t)$ is denoted as $S_{z_j}(f)$. Note that the noise $z_j(t)$ can be colored and, hence, its PSD $S_{z_j}(f)$ is generally non-white.

The RU $j$ communicates the received signal $y_j(t)$ to the CU after compression. In order to facilitate analysis and design, we follow the standard random coding approach of information theory and model the effect of quantization by means of an additive quantization noise (see, e.g., \cite{Cov06}). The compressed signal $\hat{y}_j(t)$ available at the CU is hence modeled as $\hat{y}_j(t)=y_j(t)+q_j(t)$, where the random process $q_j(t)$ is independent of $y_j(t)$ and represents the quantization noise. The quantization noise is assumed to be stationary Gaussian with zero mean and PSD $S_{q_j}(f)$. The assumption of Gaussianity is justified for its analytical tractability and by the fact that a high-dimensional dithered lattice quantizer, such as Trellis Coded Quantization \cite{Zamir96TIT, Marce90TCOM, Cal95TIT}, preceded by a linear transform can obtain a Gaussian quantization noise with any desired quantization spectrum. Note that perfect synchronization is assumed between the RUs and the CU, but a clock mismatch may be accounted for in $\hat{y}_j(t)$ by following the same approach in \cite{JSA15TVT}, which is not discussed here. Based on the quantized signals $\hat{y}_j(t)$ for all $j \in \mathcal{N}_r$, the CU estimates the target's position $\pmb{p}$. 

The selection of the PSD $S_{q_j}(f)$ is constrained by the fronthaul capacity $C_j$. Specifically, for each channel realization $g_j$, by random coding arguments \cite{Cov06}, the rate produced by the quantization operation in $\hat{y}_j(t)$ is bounded below by the mutual information $T^{-1}I(\{y_j(t)\}_{t=0}^{T};\{\hat{y}_j(t)\}_{t=0}^{T})$, where $T$ is the transmission period. Note that, throughout the paper, the mutual information is computed for a given realization of the channel gains $g_j$ for $j \in \mathcal{N}_r$. We will use the discussed information-theoretic bound in order to formulate the design problem based on the fact that quantization schemes exist that are known to operate at rates close to the information-theoretic limit \cite{Cov06}. Moreover, one could account for suboptimal quantization by modeling explicitly the gap to the information-theoretic limit.

We impose a long-term fronthaul capacity constraint. Specifically, we assume that the fronthaul capacity can be shared across multiple realizations of the target-RU channels. This happens, for instance, if the RUs can quantize the signal received across multiple coherence times of the target-RU channels. This leads to the constraint 
\begin{equation}\label{eq:fronthaulInfo}
C_j \ge \frac{1}{T}E_{g_j}\left[I\left(\left\{y_j(t)\right\}_{t=0}^{T};\left\{\hat{y}_j(t)\right\}_{t=0}^{T}\right)\right].
\end{equation}
The constraint (\ref{eq:fronthaulInfo}) has the further advantage of admitting a simple bound that can be calculated at the CU given only the available information about the average power $\sigma_{\alpha_j}^2$. To obtain such bound, we first apply Szeg$\ddot{\text{o}}$'s theorem \cite{Wyner94TIT} to (\ref{eq:fronthaulInfo}), assuming that $T$ is sufficiently large, and rewrite the constraint (\ref{eq:fronthaulInfo}) as 
\begin{eqnarray}\label{eq:fronthaul}
C_j &\ge& E_{g_j}\left[\frac{1}{B}\int_{-\frac{B}{2}}^{\frac{B}{2}}\log_2\left(1+\frac{g_j^2S_x(f)+S_{z_j}(f)}{S_{q_j}(f)}\right)df\right]\nonumber\\
&\triangleq& E_{g_j}\left[R_j(g_j, S_{q_j})\right],
\end{eqnarray}
where $B$ is the bandwidth and we have defined the function $R_j(g_j, S_{q_j})$ as the argument of the expectation in (\ref{eq:fronthaul}).
We then apply Jensen's inequality to the function $R_j(g_j, S_{q_j})$, which is concave in $g_j$, yielding the stricter constraint
\begin{eqnarray}\label{eq:rateconst}
C_j &\ge& \frac{1}{B}\int_{-\frac{B}{2}}^{\frac{B}{2}}\log_2\left(E_{g_j}\left[1+\frac{g_j^2S_x(f)+S_{z_j}(f)}{S_{q_j}(f)}\right]\right)df\nonumber\\
 &=& R_j(\sigma_{g_j}, S_{q_j}).
\end{eqnarray} 
Note that (\ref{eq:rateconst}) implies (\ref{eq:fronthaul}), and hence any solution feasible with respect to (\ref{eq:rateconst}) is also feasible with respect to (\ref{eq:fronthaul}). Via numerical results, we have verified that the bound (\ref{eq:rateconst}) is very close to the average $E_{g_j}[R_j(g_j, S_{q_j})]$ in (\ref{eq:fronthaul}) as long as the power $\sigma_{g_j}^2$ is not too large (e.g., for $\alpha_j$ following a Rayleigh fading distribution with $\sigma_{\alpha_j}^2 [dB]=10\log_{10}\sigma_{\alpha_j}^2 \le 30$ dB, distance $d_j$ larger than $200$ m and path loss exponent $\mu$ no smaller than $2$).
\subsection{Performance Metric for Localization}\label{sec:CRB}
The localization performance is measured by the squared position error (SPE) $\rho(\pmb{p}, \pmb{S}_q)=E_{\hat{\pmb{y}}, \pmb{g}}[||\hat{\pmb{p}}(\hat{\pmb{y}})-\pmb{p}||^2]$ \cite{Li13TON,Shen14ACM, JSA15TVT,Kay93Book}, where $\hat{\pmb{p}}(\hat{\pmb{y}})$ is the estimate of the target location performed at the CU based on the knowledge of the quantized received signals $\hat{\pmb{y}}=[\hat{y}_1 \cdots \hat{y}_{N_r}]^T$, with $\hat{y}_j$ being a shorthand for $\{\hat{y}_j(t)\}_{t=0}^T$; $\pmb{S}_q=[S_{q_1} \cdots S_{q_{N_r}}]^T$ collects all the PSDs of the quantization noises (suppressing the dependence on the frequency for simplicity of notation); and $\pmb{g}=[g_1 \cdots g_{N_r}]^T$. In SPE $\rho(\pmb{p}, \pmb{S}_q)$, we have made explicit the dependence on the position $\pmb{p}$ and the quantization noise PSDs $\pmb{S}_q$. Note also that the expectation in SPE $\rho(\pmb{p}, \pmb{S}_q)$ is taken over the joint distribution of the received signals and of the channel fading gains. To evaluate this quality based on the information available at the CU, we proceed as follows.
The SPE $\rho(\pmb{p}, \pmb{S}_q)$ is first bounded by the CRB, i.e., $\rho(\pmb{p}, \pmb{S}_q) \ge E_{\pmb{g}}[\text{tr}\{\pmb{J}^{-1}(\pmb{p}, \pmb{g}, \pmb{S}_q)\}]$, where $\pmb{J}(\pmb{p}, \pmb{g}, \pmb{S}_q)$ is the Equivalent Fisher Information Matrix (EFIM) for the estimation of the target's position $\pmb{p}$ (see, e.g., \cite{JSA15TVT, Li13TON,Shen14ACM, Kay93Book}). We recall that the need to resort to the EFIM stems from the presence of the unknown parameter $\pmb{g}$ \cite{JSA15TVT, Shen14ACM, Li13TON}. In light of this bound, and given its analytical tractability, we will use the CRB as the performance metric for localization. Similar to \cite{JSA15TVT}, the EFIM for the position of target is calculated as 
\begin{eqnarray}\label{eq:EFIM}
&&\hspace{-1cm}\pmb{J}(\pmb{p}, \pmb{g}, \pmb{S}_q)\nonumber\\
&&\hspace{-1cm}=\sum_{j \in \mathcal{N}_r}\pmb{J}_\phi(\phi_j)\frac{8\pi^2g_j^2}{c^2}\int_{-\infty}^{\infty}\frac{f^2S_x(f)}{S_{z_j}(f)+S_{q_j}(f)}df,
\end{eqnarray}
where we have defined the direction matrix $\pmb{J}_\phi(\phi)=[\cos^2(\phi) \,\,\, \cos(\phi)\sin(\phi);
\cos(\phi)\sin(\phi) \,\,\, \sin^2(\phi) ]$. In the following, given EFIM $\pmb{J}(\pmb{p}, \pmb{g}, \pmb{S}_q)$, we will also use the notation $\pmb{J}(\pmb{\phi}, \pmb{g}, \pmb{S}_q)$ for the EFIM to emphasize the dependence on the the inter-node angles $\pmb{\phi}=[\phi_1\cdots \phi_{N_r}]^T$. In order to deal with the expectation in the CRB, $E_{\pmb{g}}[\text{tr}\{\pmb{J}^{-1}(\pmb{p}, \pmb{g}, \pmb{S}_q)\}]$, over the channel gains $\pmb{g}$, we again leverage the Jensen's inequality as in (\ref{eq:rateconst}) to obtain the inequality
\begin{eqnarray}\label{eq:crbJen}
&&\hspace{-0.8cm}E_{\pmb{g}}\hspace{-0.05cm}\left[\text{tr}\hspace{-0.05cm}\left\{\hspace{-0.05cm}\pmb{J}^{-1}\hspace{-0.05cm}\left(\pmb{p}, \pmb{g}, \pmb{S}_q\hspace{-0.05cm}\right)\right\}\hspace{-0.05cm}\right]\nonumber\\
&&\hspace{-0.8cm} \ge\hspace{-0.05cm} \text{tr}\hspace{-0.1cm}\left\{\hspace{-0.1cm}\left(\hspace{-0.15cm}E_{\pmb{g}}\hspace{-0.15cm}\left[\sum_{j \in \mathcal{N}_r}\hspace{-0.1cm}\pmb{J}_\phi(\phi_j)\frac{8\pi^2g_j^2}{c^2}\hspace{-0.1cm}\int_{-\infty}^{\infty}\hspace{-0.1cm}\frac{f^2S_x(f)}{S_{z_j}(f)+S_{q_j}(f)}df\hspace{-0.1cm}\right]\hspace{-0.05cm}\right)^{\hspace{-0.05cm}-1}\hspace{-0.1cm}\right\}
\nonumber\\
&&\hspace{-0.8cm}=\hspace{-0.05cm} \text{tr}\hspace{-0.05cm}\left\{\hspace{-0.05cm}\pmb{J}^{-1}\hspace{-0.1cm}\left(\pmb{p}, \pmb{\sigma}_{g}, \pmb{S}_q\hspace{-0.05cm}\right)\hspace{-0.05cm}\right\}\hspace{-0.1cm},
\end{eqnarray}
where $\pmb{\sigma}_g=[\sigma_{g_1}\cdots\sigma_{g_{N_r}}]^T$. In the following, similar to the discussion around the fronthaul constraint (\ref{eq:rateconst}), we will adopt the lower bound in (\ref{eq:crbJen}) as the performance metric for the localization accuracy.  
\subsection{Problem Formulation}\label{sec:prob}
\begin{figure}[t]
\begin{center}
\includegraphics[width=9cm]{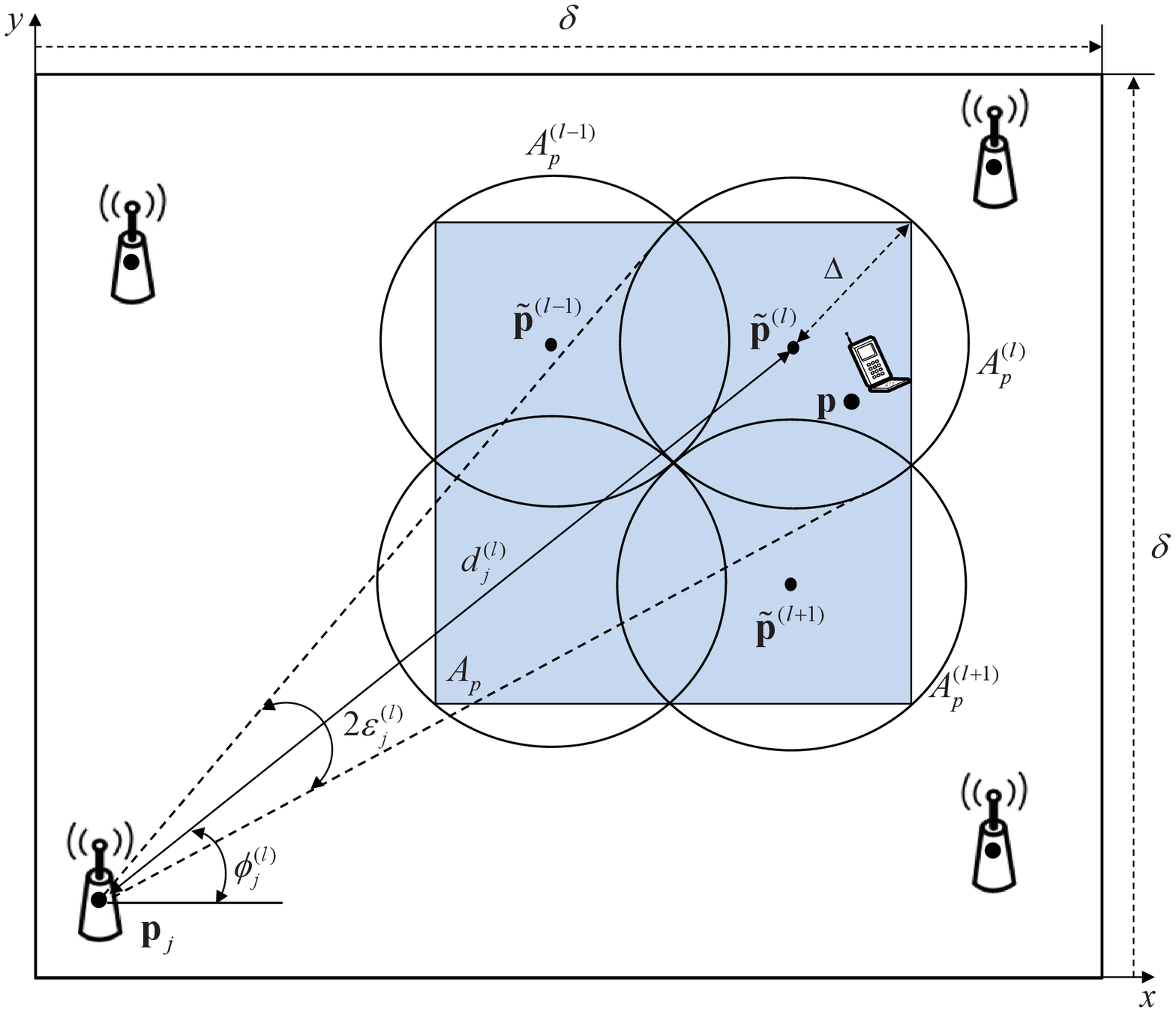}
\caption{The target is in the shaded area $A_p$. The region $A_p$ is covered by a set of circular areas $\{A_p^{(l)}\}_{l \in \mathcal{L}}$, with radius $\Delta$, each of which is centered at location $\tilde{\pmb{p}}^{(l)}$.} \label{fig:uncer}
\end{center}
\end{figure}   
Here, we formulate the problem of optimizing the quantization strategy under fronthaul capacity constraints. As mentioned, the a priori information available at the CU about the position of the target is characterized by the uncertainty area $A_p$, which is generally included in the overall square region. An example is illustrated in Fig. \ref{fig:uncer}. The region $A_p$ is arbitrary but is assumed to exclude the regions very close to the RUs in order to ensure the validity of the path loss model. To simplify the analysis, the region $A_p$ is described, as in \cite{Shen14ACM}, as the union of a finite set of circular areas $\{A_p^{(l)}\}_{l \in \mathcal{L}}$, in the sense that we have the inclusion relationship $A_p \subseteq \cup_{l \in \mathcal{L}}A_p^{(l)}$. Each circle $A_p^{(l)}$ is centered at $\tilde{\pmb{p}}^{(l)}$ and has radius $\Delta$. Note that a larger radius $\Delta$ generally leads to a less accurate approximation of the uncertainty region $A_p$, but, as it will be seen below, it reduces the complexity of the resulting algorithm. As shown in Fig. \ref{fig:uncer}, when the circle $A_p^{(l)}$ includes the target $\pmb{p}$, i.e., $\pmb{p} \in A_p^{(l)}$, the actual inter-node distances and angles lie in uncertainty sets $d_j \in S_{d_j}^{(l)} \triangleq [d_j^{(l)}-\Delta, d_j^{(l)}+\Delta]$ and $\phi_j \in S_{\phi_j}^{(l)} \triangleq [\phi_j^{(l)}-\epsilon_j^{(l)}, \phi_j^{(l)}+\epsilon_j^{(l)}]$ for all $j \in \mathcal{N}_r$, respectively, where $\epsilon_j^{(l)}=\arcsin(\Delta/d_j^{(l)})$ is the angular uncertainty defined by the radius $\Delta$ \cite{Shen14ACM}. Moreover, from the definition of uncertainty sets and of $g_j$, the average channel power gain lies in the interval $\sigma_{g_j} \in S_{\sigma_{g_j}}^{(l)} \triangleq [\sigma_{L,g_j}^{(l)}, \sigma_{U,g_j}^{(l)}]$, where $\sigma_{L,g_j}^{(l)}=\sigma_{\alpha_j}/(d_j^{(l)}+\Delta)^\mu$ and $\sigma_{U,g_j}^{(l)}=\sigma_{\alpha_j}/(d_j^{(l)}-\Delta)^\mu$. Considering the union over all the circular regions, the angular positions and the average channel power gains lie in the uncertainty set $\mathcal{U}$ such as $(\pmb{\phi}, \pmb{\sigma}_g) \in \mathcal{U}=\bigcup_{l \in \mathcal{L}}\mathcal{U}_l$, where $\mathcal{U}_l=\prod_{j \in \mathcal{N}_r} S_{\phi_j}^{(l)} \times S_{\sigma_{g_j}}^{(l)}$. The uncertainty set $\mathcal{U}$ is assumed to be a priori information available for the optimization.

Following the robust optimization methodology introduced in \cite{Ben09Book}, we wish to optimize the PSDs of quantization noises with the aim of minimizing the worst-case localization error of the target. Specifically, we are solving the problem  
\begin{subequations}\label{eq:Ropt}
\begin{eqnarray}
&&\hspace{-0.7cm}{\mathop {\text{min} }\limits_{{S_{q_j}(f)\ge 0}}}\hspace{0.1cm} {\mathop {\text{max} }\limits_{{l \in \mathcal{L}}}}\hspace{0.1cm}{\mathop {\text{max} }\limits_{{(\pmb{\phi}, \pmb{\sigma}_g) \in \mathcal{U}_l}}}\hspace{0.5cm}\text{tr}\left\{\pmb{J}^{-1}(\pmb{\phi}, \pmb{\sigma}_g, \pmb{S}_q)\right\} \label{eq:Roptobj}  \\
&& \hspace{-0.3cm}{\rm{s.t.}}\hspace{0.1cm} {\mathop {\text{max} }\limits_{{\sigma_{g_j}' \in \bigcup_{l' \in \mathcal{L}}S_{\sigma_{g_j}}^{(l')}}}}{\left(R_j(\sigma_{g_j}',S_{q_j}) -C_j\right) \le  0,\hspace{0.1cm}\text{for}\hspace{0.1cm}j \in \mathcal{N}_r}\label{eq:Roptconst}.
 \end{eqnarray}
\end{subequations} 
Note that the constraint (\ref{eq:Roptconst}) guarantees the feasibility of the solution with respect to the fronthaul constraint no matter what the channel gain is, and hence irrespective of the target distance within the uncertainty region. Note also that, since the worst-case values of $\sigma_{g_j}$, for the objective (\ref{eq:Roptobj}) and the constraint (\ref{eq:Roptconst}) need not be the same, we differentiate $\sigma_{g_j}$ from $\sigma_{g_j}'$.  
\section{Optimization of Fronthaul Quantization}\label{sec:opt}
In this section, we propose an algorithm that aims at minimizing the worst-case SPE under fronthaul capacity constraints over the fronthaul quantization noise PSDs as per problem (\ref{eq:Ropt}). To this end, we first address the inner optimization problems over $(\pmb{\phi}, \pmb{\sigma}_g) \in \mathcal{U}_l$ and $\sigma_{g_j}' \in \bigcup_{l' \in \mathcal{L}}S_{\sigma_{g_j}'}^{(l')}$ in Sec. \ref{sec:optCircle}, and then consider the outer optimizations over $l$ and $\pmb{S}_q$ in Sec. \ref{sec:optLandSq}. The proposed fronthaul quantization design is summarized in Algorithm $1$.
\subsection{Optimization over $(\pmb{\phi}, \pmb{\sigma}_g)$ and $\sigma_{g_j}'$}\label{sec:optCircle}
We here focus on the optimizations over $(\phi,\sigma_g)$ for the maximal CRB and over $\sigma_{g_j}'$ for the maximal rate within $\mathcal{U}_l$ in (\ref{eq:Ropt}), namely $\max_{(\pmb{\phi}, \pmb{\sigma}_g) \in \mathcal{U}_l}\text{tr}\{\pmb{J}^{-1}(\pmb{\phi}, \pmb{\sigma}_g, \pmb{S}_q)\}$ and $\max_{\sigma_{g_j}' \in \bigcup_{l' \in \mathcal{L}}S_{\sigma_{g_j}}^{(l')}}(R_j(\sigma_{g_j}',S_{q_j}) -C_j)$ for given PSDs $\pmb{S}_q$. Both functions $\text{tr}\{\pmb{J}^{-1}(\pmb{\phi}, \pmb{\sigma}_g, \pmb{S}_q)\}$ and $R_j(\sigma_{g_j}', S_{q_j})$ are monotonically non-increasing and non-decreasing functions of the channel gain standard deviations $\sigma_{g_j}$, respectively. This leads us immediately to conclude that the maximizations at hand are achieved at the smallest possible value for $\max_{(\pmb{\phi}, \pmb{\sigma}_g) \in \mathcal{U}_l}\text{tr}\{\pmb{J}^{-1}(\pmb{\phi}, \pmb{\sigma}_g, \pmb{S}_q)\}$ and the largest value for $\max_{\sigma_{g_j}' \in \bigcup_{l' \in \mathcal{L}}S_{\sigma_{g_j}}^{(l')}}(R_j(\sigma_{g_j}',S_{q_j}) -C_j)$, namely, respectively, at $\sigma_{g_j}=\sigma_{L,g_j}^{(l)}$ and $\sigma_{g_j}'=\sigma_{U,g_j}^{(l')}$, irrespective of the values of $\pmb{\phi}$ and $\pmb{S}_q$ for both problems. The maximization over the angle $\pmb{\phi}$ in $\max_{(\pmb{\phi}, \pmb{\sigma}_g) \in \mathcal{U}_l}\text{tr}\{\pmb{J}^{-1}(\pmb{\phi}, \pmb{\sigma}_g, \pmb{S}_q)\}$ is instead carried out by following the relaxation method introduced in \cite{Li13TON}. To this end, let us define, for every circle $l \in \mathcal{L}$, the matrix $\pmb{Q}_\phi(\phi_j^{(l)})$ as $\pmb{Q}_\phi(\phi_j^{(l)})=\pmb{J}_\phi(\phi_j^{(l)})-\sin\epsilon_j^{(l)}\pmb{I}$. We also define the matrix $\pmb{Q}(\pmb{\sigma}_g, \pmb{S}_q)=\sum_{j \in \mathcal{N}_r}\pmb{Q}_\phi(\phi_j^{(l)})8\pi^2\sigma_{g_j}^2/c^2\int_{-\infty}^{\infty}f^2S_x(f)/(S_{z_j}(f)+S_{q_j}(f))df$, which is obtained by using $\pmb{Q}_\phi(\phi_j^{(l)})$ in lieu of $\pmb{J}_\phi(\phi)$ in (\ref{eq:EFIM}).  
Then, if $\pmb{Q}(\pmb{\sigma}_g, \pmb{S}_q)\succeq 0$, denoting $\pmb{\sigma}_{L,g}^{(l)}=[\sigma_{L,g_1}^{(l)} \cdots \sigma_{L,g_{N_r}}^{(l)}]^T$, the worst-case CRB $\max_{(\pmb{\phi}, \pmb{\sigma}_g) \in \mathcal{U}_l}\text{tr}\{\pmb{J}^{-1}(\pmb{\phi}, \pmb{\sigma}_g, \pmb{S}_q)\}$ is upper bounded as ${\mathop {\text{max} }\limits_{{(\pmb{\phi}, \pmb{\sigma}_g) \in \mathcal{U}_l}}}\text{tr}\{\pmb{J}^{-1}(\pmb{\phi}, \pmb{\sigma}_g, \pmb{S}_q)\} = \max_{\pmb{\phi} \in \prod_{j \in \mathcal{N}_r} S_{\phi_j}^{(l)}}$ $\text{tr}\{\pmb{J}^{-1}(\pmb{\phi}, \pmb{\sigma}_{L,g}^{(l)}, \pmb{S}_q)\} \le \text{tr}\{\pmb{Q}^{-1}(\pmb{\sigma}_{L,g}^{(l)}, \pmb{S}_q)\}$, where the first equality follows from the discussion above and the second inequality is as in \cite{Li13TON, JSA15TVT}. This inequality provides a conservative measure of the worst-case CRB for all positions within the circle $A_p^{(l)}$. We will adopt the bound $\text{tr}\{\pmb{Q}^{-1}(\pmb{\sigma}_{L,g}^{(l)}, \pmb{S}_q)\}$ as the performance criterion, and the validity of this choice will be validated in Section \ref{sec:numerical} by elaborating on the performance of the proposed algorithm via numerical results. 
\subsection{Optimization over $l$ and $\pmb{S}_q$}\label{sec:optLandSq}
Given the discussion above, the optimization problem (\ref{eq:Ropt}) is restated in the more conservative formulation
\begin{subequations}\label{eq:RRopt}
\begin{eqnarray}
&&\hspace{-0.8cm}{\mathop {\text{min} }\limits_{{S_{q_j}(f) \ge 0}}} \hspace{0.1cm}{\mathop {\text{max} }\limits_{{l \in \mathcal{L}}}}\hspace{1.1cm}{\text{tr}\left\{\pmb{Q}^{-1}(\pmb{\sigma}_{L,g}^{(l)}, \pmb{S}_q)\right\}}\label{eq:RRoptobj}  \\
&& \hspace{-0.3cm}{\rm{s.t.}}\hspace{0.5cm} {R_j(\sigma_{U,g_j}^{(l')}, S_{q_j})\le C_j, \hspace{0.1cm}\text{for}\hspace{0.1cm}l' \in \mathcal{L}\hspace{0.1cm}\text{and}\hspace{0.1cm}j \in \mathcal{N}_r}\label{eq:RRoptconst}.
 \end{eqnarray}
\end{subequations}
In order to address problem (\ref{eq:RRopt}), we first make the change of variables $A_{q_j}(f)=1/S_{q_j}(f)$. This is done in order to avoid the unbounded solution $S_{q_j}(f)=\infty$ for frequencies that are neglected by the quantization and hence have infinite quantization noise. Note that, due to (\ref{eq:RRoptconst}), the solution $S_{q_j}(f)=0$, and hence $A_{q_j}(f)=\infty$ is not feasible for any finite $C_j$. Then, we consider the epigraph formulation of (\ref{eq:RRopt}) which is given as 
\begin{subequations}\label{eq:epiopt}
\begin{eqnarray}
&&\hspace{-1.4cm}{\mathop {\text{min} }\limits_{{A_{q_j}(f) \ge 0},t}} \hspace{0.5cm}{t}\\
&& \hspace{-0.8cm}{\rm{s.t.}}\hspace{0.5cm} {\text{tr}\left\{\pmb{Q}^{-1}\left(\pmb{\sigma}_{L,g}^{(l)}, \pmb{A}_q\right)\right\} -t \le 0, \hspace{0.1cm}\text{for}\hspace{0.1cm}l \in \mathcal{L},}\label{eq:epiCRBconst}\\
&& \hspace{0.1cm} {R_j(\sigma_{U,g_j}^{(l)}, A_{q_j})\le C_j, \hspace{0.1cm}\text{for}\hspace{0.1cm}l \in \mathcal{L} \hspace{0.1cm}\text{and}\hspace{0.1cm}j\in \mathcal{N}_r}\label{eq:epifronthaulconst},
\end{eqnarray}
\end{subequations}
where $\pmb{A}_q=[A_{q_1} \cdots A_{q_{N_r}}]^T$. Note that in the epigraph formulation (\ref{eq:epiopt}), there is no need to distinguish between $l$ and $l'$ as in (\ref{eq:RRopt}). Using the Charnes-Cooper transformation \cite{Charnes62Naval}, $m_j(f)=A_{q_j}(f)/(1+S_{z_j}(f)A_{q_j}(f))$ and $n_j(f)=1/(1+S_{z_j}(f)A_{q_j}(f))$, the problem (\ref{eq:epiopt}) can be transformed to the equivalent problem in (\ref{eq:Topt}), 
\begin{subequations}\label{eq:Topt}
\begin{eqnarray}
&&\hspace{-2.8cm}{\mathop {\text{min} }\limits_{0 \le m_j(f), 0 \le n_j(f) <1, t}} \hspace{0.2cm}\hspace{1.1cm}{t}\label{eq:Toptobj}\\
&& \hspace{-1.5cm}{\rm{s.t.}}\hspace{0.7cm} {\text{tr}\left\{\left(\sum_{j \in \mathcal{N}_r}\pmb{Q}_\phi(\phi_j^{(l)})\frac{8\pi^2\sigma_{L,g_j}^{(l)2}}{c^2}\int_{-\infty}^{\infty}f^2S_x(f)m_j(f)df\right)^{-1}\right\} -t \le 0, \hspace{0.5cm}\text{for}\hspace{0.2cm}l \in \mathcal{L},}\label{eq:TCRBconst}\\
&& \hspace{-0.3cm} {\frac{1}{B}\int_{-\frac{B}{2}}^{\frac{B}{2}}\log_2\left(\frac{1+\sigma_{U,g_j}^{(l)2}S_x(f)m_j(f)}{n_j(f)}\right)df\le C_j, \hspace{0.5cm}\text{for}\hspace{0.2cm}l \in \mathcal{L} \hspace{0.2cm}\text{and}\hspace{0.2cm}\text{for}\hspace{0.2cm}j\in \mathcal{N}_r}\label{eq:Tfronthaulconst},\\
&& \hspace{-0.3cm} {S_{z_j}(f)m_j(f)+n_j(f)=1, \hspace{0.5cm}\text{for}\hspace{0.2cm}j \in \mathcal{N}_r,}
\end{eqnarray}
\end{subequations}
where $\pmb{m}=[m_1 \cdots m_{N_r}]^T$, $\pmb{n}=[n_1 \cdots n_{N_r}]^T$ and $m_j$ and $n_j$ are shorthands for the functions $m_j(f)$ and $n_j(f)$, respectively. Note that the number of constraints (\ref{eq:TCRBconst}) and (\ref{eq:Tfronthaulconst}) depends on the number of circular areas used to approximate the uncertainty region $A_p$ (see Fig. \ref{fig:uncer}). The optimization problem (\ref{eq:Topt}) is complicated since (\textit{i}) the unknowns $m_j(f)$ and $n_j(f)$ are continuous functions of the frequency $f$; (\textit{ii}) the constraint (\ref{eq:Tfronthaulconst}) is not convex. Note that the objective function (\ref{eq:Toptobj}) is convex. To deal with (\textit{i}), we discretize the frequency domain using a uniform quantization of the frequency axis with $N_f$ equally spaced points. As for (\textit{ii}), we leverage the standard DC method \cite{Tao05Springer}. In particular, (\ref{eq:Tfronthaulconst}) can be written as a DC functions and hence a locally tight upper bound can be obtained by linearizing the negative convex function. The details are shown in Algorithm $1$. We note that each iteration of the DC algorithm provides a feasible solution and that the sequence of objective functions is non-increasing \cite{Tao05Springer}.
\section{Numerical Results}\label{sec:numerical}
In this section, we evaluate the performance of the proposed algorithm. For reference, we consider a baseline scheme that does not optimize the PSDs and instead assumes white quantization noise PSDs $S_{q_j}(f)=\sigma_{q_j}^2$, where $\sigma_{q_j}^2$ is computed by imposing equality in the constraint (\ref{eq:RRoptconst}) for the circle $l'$ having the maximal rate among $l' \in \mathcal{L}$. 
\begin{figure}[t]
\begin{center}
\includegraphics[width=12cm]{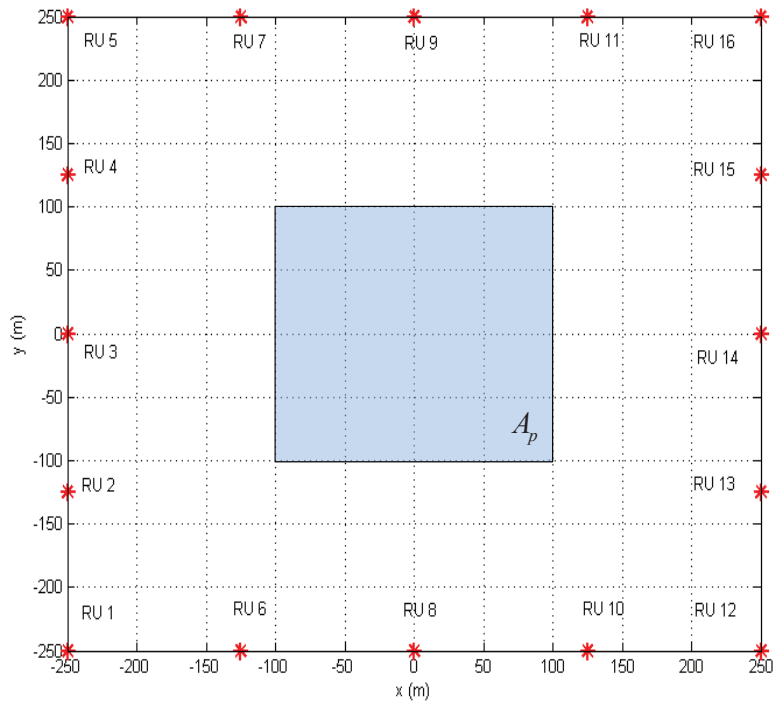}
\caption{Topology for the examples discussed in Section \ref{sec:opt}.} \label{fig:RU}
\end{center}
\end{figure} 

In the following numerical results, the size of the area is $\delta=500$ m and the path loss exponent is $\mu=3$. Moreover, the two-sided bandwidth is $B=1$ MHz and the channel power is normalized to $\sigma_{\alpha_j}^2=1$ for all $j \in \mathcal{N}_r$. The ESD of the signal is $S_x(f)=-60$ dBm/Hz in the bandwidth $[-B/2, B/2]$, and $S_x(f)=0$, otherwise. Accounting for both thermal noise and interference, the channel noises $z_j(t)$ for all $j \in \mathcal{N}_r$ are assumed to follow a standard autoregressive model of order $1$ with correlation coefficient $\rho$, so that the noise PSDs are $S_z(f)=N_0(1-\rho^2)/(\left|1-\rho e^{-j2\pi f/B}\right|^2)$ with parameters $N_0=-174$ dBm/Hz and $\rho=0.9$. We assume the rectangular uncertainty region $A_p$ is a square, of size $200$ m $\times 200$ m and is centered as shown in Fig. \ref{fig:RU}. We set the radius of the circular areas $\{A_p^{(l)}\}_{l \in \mathcal{L}}$ that cover the uncertainty regions to $\Delta=50\sqrt{2}$ m and choose the centers $\tilde{\pmb{p}}^{(l)}$ as $\tilde{\pmb{p}}^{(1)}=[-50 \,\,\, -50]^T$, $\tilde{\pmb{p}}^{(2)}=[-50 \,\,\, 50]^T$, $\tilde{\pmb{p}}^{(3)}=[50 \,\,\, -50]^T$ and $\tilde{\pmb{p}}^{(4)}=[50 \,\,\, 50]^T$ so that the number of circular regions is $|\mathcal{L}|=4$. We consider $N_r=16$ RUs, equally spaced along each side as illustrated in Fig. \ref{fig:RU} and impose an equal fronthaul capacity constraint $C_j=C$ for all RUs $j \in \mathcal{N}_r$. Furthermore, we set $N_f=100$ for discretizing the frequency axis. 

\begin{figure}[t]
\begin{center}
\includegraphics[width=12cm]{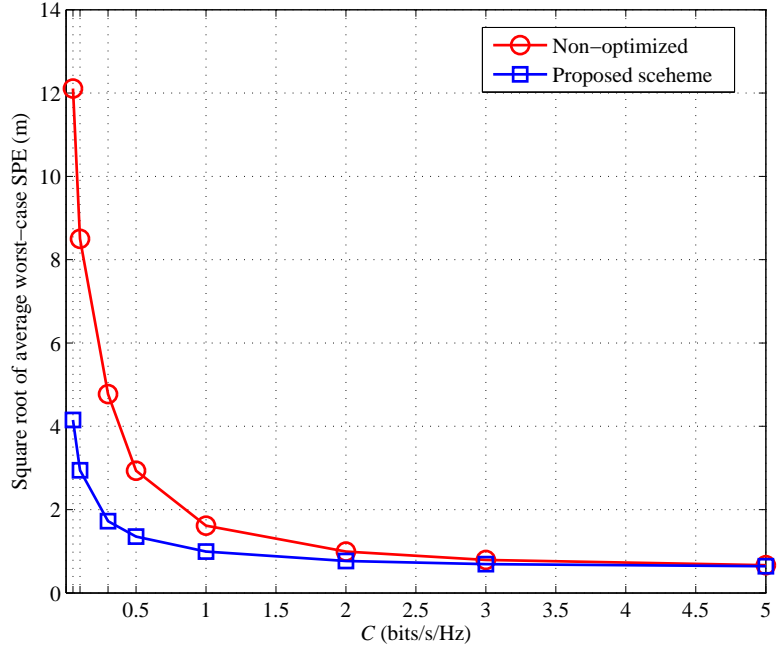}
\caption{Square root of average worst-case SPE as a function of the fronthaul capacity $C$ for the set-up in Fig. \ref{fig:RU}.} \label{fig:CSq}
\end{center}
\end{figure}
Fig. \ref{fig:CSq} shows the square root of the average worst-case SPE as a function of the fronthaul capacity $C$. To evaluate this quantity, we generated $400$ target's positions uniformly distributed in uncertainty region $A_p$. For each position, we calculated the CRB by using the PSDs $S_{q_j}(f)$ for $j \in \mathcal{N}_r$ obtained via Algorithm $1$ or by using the baseline white PSDs. Specifically, by means of Monte Carlo simulation, we evaluated the average CRB with respect to the channel fading coefficients $\alpha_j$ for $j \in \mathcal{N}_r$, which are independent and follow the Rayleigh distribution with unit power. Finally, we chose the average worst SPE for the given $C$ across all considered positions and computed its square root. In Fig. \ref{fig:CSq}, a larger fronthaul capacity $C$ yields an improved localization. It is also observed that the proposed design outperforms the baseline non-optimized solution at low-to-moderate values of $C$. For instance, for $C = 0.1$ bits/s/Hz, the proposed scheme obtains a square root of average worst-case SPE of around 2.9 m, while the non-optimized strategy provides a localization error of around 8.5 m. We observe that, in comparison to the fronthaul rates needed to support data communication, see, e.g., \cite{Simeone11FNT}, localization has lower requirements. This is not surprising since localization only requires the CU to estimate the target's position and not a data stream. 

\begin{figure}[t]
\begin{center}
\includegraphics[width=12cm]{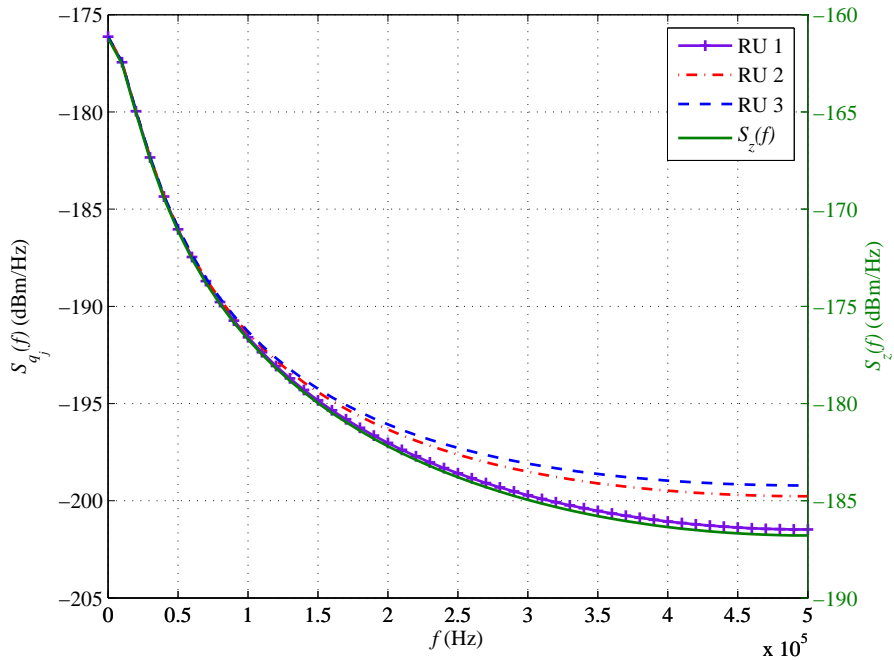}
\caption{Quantization noise $S_{q_j}(f)$ obtained by the proposed algorithm for RUs $1$, $2$ and $3$ along with the noise PSDs $S_z(f)$ for $C=5$ bits/s/Hz.} \label{fig:Sqdesign}
\end{center}
\end{figure}
In Fig. \ref{fig:Sqdesign}, we show the quantization noise PSDs $S_{q_j}(f)$ obtained when the fronthaul capacity constraint is $C=5$ bits/s/Hz. Due to the symmetric topology of the RUs, we only show the PSDs for the RU $1$, RU $2$ and RU $3$. As shown in Fig. \ref{fig:Sqdesign}, a larger quantization noise is assigned by all RUs to frequencies at which the noise PSD is more pronounced, hence compressing more accurately at frequencies that are less affected by noise. 
\section{Concluding Remarks}\label{sec:con} 
In this paper, we have investigated fronthaul quantization design for Cloud Radio Positioning based on direct localization. Under the assumption of synchronous RUs, a robust, or worst-case, optimization formulation is adopted and the resulting algorithm is verified via numerical results. Among the open issues left for future work, we mention here the comparison between direct and indirect localization as a function of the available fronthaul capacity, extending \cite{Weiss04TSP}, and the optimization of fronthaul quantization in the presence of a timing mismatch between RUs and CU.   
\begin{figure}[t]
\begin{flushleft}
\begin{tabular}{|p{18cm}|}
\hline
$\bf{Algorithm}$ $\bf{1}$: Robust fronthaul quantization algorithm (problem (\ref{eq:Topt}))\\
\hline
1. Initialize a nonnegative $\pmb{m}^{(1)}$ and a sufficiently large positive integer $N_f$.\\ 
2. (\textit{DC algorithm}) Update $\pmb{m}^{(i+1)}$ as a solution of the following convex problem:
\begin{subequations}\label{eq:ToptMM}
\begin{eqnarray}
&&\hspace{-1.8cm}{\mathop {\text{min} }\limits_{\pmb{m}^{(i+1)},\pmb{n},t}} \hspace{0.5cm}{t}\\
&& \hspace{-1.4cm}{\rm{s.t.}}\hspace{0.6cm} \text{tr}\left\{\left(\sum_{j=1}^{N_r}\pmb{Q}_\phi(\phi_j^{(l)})\frac{8\pi^2\sigma_{L,g_j}^{(l)2}}{c^2}\sum_{n=1}^{N_f/2}\frac{2B}{N_f}f_n^2S_x(f_n)m_j^{(i+1)}(f_n) \right)^{-1}\right\} -t \le 0 \hspace{0.1cm}\text{for}\hspace{0.1cm}l \in \mathcal{L},\label{eq:TCRBconstMM}\\
&& \hspace{-0.3cm} \sum_{n=1}^{N_f/2}\frac{2}{N_f}\left[h\left(m_j^{(i+1)}(f_n),m_j^{(i)}(f_n)\right)-\log_2\left(n_j(f_n)\right)\right]\le C_j\hspace{0.1cm}\text{for}\hspace{0.1cm}l \in \mathcal{L} \hspace{0.1cm}\text{and}\hspace{0.1cm}j\in\mathcal{N}_r\label{eq:TfronthaulconstMM},\\
&& \hspace{-0.3cm} \sum_{j=1}^{N_r}\pmb{Q}_\phi(\phi_j^{(l)})\frac{8\pi^2\sigma_{L,g_j}^{(l)2}}{c^2}\sum_{n=1}^{N_f/2}\frac{2B}{N_f}f_n^2S_x(f_n)m_j^{(i+1)}(f_n)\succeq 0\hspace{0.1cm}\text{for}\hspace{0.1cm}l \in \mathcal{L},\\
&& \hspace{-0.3cm} {S_{z_j}(f_n)m_j^{(i+1)}(f_n)+n_j(f_n)=1\hspace{0.1cm}\text{for}\hspace{0.1cm}j \in \mathcal{N}_r,}\\
&& \hspace{-0.3cm} {0 \le n_j(f_n) < 1\hspace{0.1cm}\text{for}\hspace{0.1cm}j \in \mathcal{N}_r,}\\
&& \hspace{-0.3cm} {m_j^{(i+1)}(f_n) \ge 0, \hspace{0.1cm}\text{for}\hspace{0.1cm}j \in \mathcal{N}_r,}
\end{eqnarray}
\end{subequations}
where $f_n=\frac{nB}{N_f}$ and $h(m_j^{(i+1)}(f_n),m_j^{(i)}(f_n))$ is the linear function defined as
\begin{eqnarray}
&&\hspace{-2.5cm}h\left(m_j^{(i+1)}(f_n),m_j^{(i)}(f_n)\right)=\log_2\left(1+\sigma_{U,g_j}^{(l)2}S_x(f_n)m_j^{(i)}(f_n)\right)\nonumber\\
&&\hspace{2.5cm}+\frac{\sigma_{U,g_j}^{(l)2}S_x(f_n)}{\ln 2 \left(1+\sigma_{U,g_j}^{(l)2}S_x(f_n)m_j^{(i)}(f_n)\right)}\times\left(m_j^{(i+1)}(f_n)-m_j^{(i)}(f_n)\right).
\end{eqnarray}
3. Stop if $\sum_{j=1}^{N_r}\sum_{n=1}^{N_f}\left\|m_j^{(i+1)}(f_n)-m_j^{(i)}(f_n)\right\|_F < \delta_{\rm{th}}$ with a predefined threshold value $\delta_{\rm{th}}$. Otherwise, $i \leftarrow i+1$ and go back to step 2.\\
4. Obtain $\pmb{S}_q$ by calculating 
\begin{equation}
S_{q_j}(f) = \left\{\frac{n_j(f_n)}{m_j^{(i+1)}(f_n)}\right\}_{n=1, \dots, N_f}.
\end{equation}
\\
\hline
\end{tabular}
\end{flushleft}
\end{figure}
\bibliographystyle{IEEEtran}
\bibliography{JSAref}

\end{document}